\documentclass[12pt,preprint,natbib]{aastex}

\RequirePackage{lineno} 
\usepackage{longtable}
\usepackage{epstopdf}  
\usepackage{ifthen}
\usepackage{rotating}
\usepackage{subfigure} 
\usepackage{booktabs}

\begin{document}
\bibliographystyle{apj}

\setlength{\footskip}{0pt} 

\title{Preliminary Analysis of {\it SOHO/STEREO} Observations of Sungrazing Comet ISON (C/2012 S1) Around Perihelion}

\author{Matthew M. Knight\altaffilmark{1,2,3}, {Karl Battams\altaffilmark{4}}}

\author{Submitted to {\it The Astrophysical Journal Letters}, 2014 Jan 2; Revised 2014 Jan 23}

\author{Manuscript: \pageref{lastpage} pages text (single spaced), \ref{lasttable} table, \ref{lastfig} figures}

\altaffiltext{1}{Contacting author: knight@lowell.edu.}
\altaffiltext{2}{Lowell Observatory, 1400 W. Mars Hill Rd, Flagstaff, Arizona 86001, USA}
\altaffiltext{3}{Visiting scientist at The Johns Hopkins University Applied Physics Laboratory, 11100 Johns Hopkins Road, Laurel, Maryland 20723, USA}
\altaffiltext{4}{Naval Research Laboratory - Code 7685, 4555 Overlook Avenue, SW, Washington, DC 20375}

\begin{singlespace}

\begin{abstract}
We present photometric and morphological analysis of the behavior of sungrazing comet C/2012 S1 ISON in {\it SOHO} and {\it STEREO} images around its perihelion on 2013 November 28.779 UT. ISON brightened gradually November 20--26 with a superimposed outburst on November 21.3--23.5. The slope of brightening changed about November 26.7 and was significantly steeper in {\it SOHO}'s orange and clear filter images until November 27.9 when it began to flatten out, reaching a peak about November 28.1 (r$_\mathrm{H}{\approx}17 R_\odot$), then fading before brightening again from November 28.6 (r$_\mathrm{H}{\approx}5 R_\odot$) until disappearing behind the occulting disc. ISON brightened continuously as it approached perihelion while visible in all other telescopes/filters. The central condensation disappeared about November 28.5 and the leading edge became progressively more elongated until perihelion. These photometric and morphological behaviors are reminiscent of the tens of meter sized Kreutz comets regularly observed by {\it SOHO} and {\it STEREO} and strongly suggest that the nucleus of ISON was destroyed prior to perihelion. 
This is much too small to support published gas production rates and implies significant mass loss and/or disruption in the days and weeks leading up to perihelion. No central condensation was seen post-perihelion. The post-perihelion lightcurve was nearly identical in all telescopes/filters and fell slightly steeper than $r_\mathrm{H}^{-2}$. This implies that the brightness was dominated by reflected solar continuum off of remnant dust in the coma/tail and that any remaining active nucleus was $<$10 m in radius.
\end{abstract}


\section{INTRODUCTION}
As comet ISON (C/2012 S1) neared its sungrazing perihelion at 0.0124 AU (2.7 solar radii, $R_\odot$) on 2013 November 28.779 UT, monitoring by ground-based optical observers became challenging and, finally, impossible. While observations continued from a handful of ground-based telescopes capable of pointing very close to the Sun (cf. \citealt{cbet3720b}, \citealt{iauc9266}, \citealt{cbet3719}), round-the-clock monitoring by observers across the globe ceased. Instead, the task of monitoring ISON fell to the telescopes on {\it Solar and Heliospheric Observatory (SOHO)} and {\it Solar TErrestrial RElations Observatory (STEREO)}, whose unobstructed views of the near-Sun environment allowed continuous observations of ISON from multiple vantage points. 


We present here ISON's {\it SOHO} and {\it STEREO} lightcurves leading up to and shortly after perihelion. This is only a subset of the full dataset obtained by these spacecraft and includes the photometric measurements most suited to rapid analysis. The data we present begin early enough to overlap with brightness and production rate measurements from the ground and extend until the post-perihelion remnants became too diffuse to allow reliable photometric measurements. We conduct some preliminary analyses, but have deliberately reserved detailed investigations for subsequent papers in order to make these data available to the community quickly.


\section{OBSERVATIONS AND REDUCTIONS}
\label{sec:observations}

\subsection{Instrumentation}
{\it SOHO} continuously observes the Sun from L1
\citep{domingo95}. It carries a suite of instruments but here we discuss only the coronagraphs ``C2'' and ``C3,'' part of the Large Angle and Spectrometric Coronagraph \citep{brueckner95}. {\it STEREO} consists of twin spacecraft in orbits near 1 AU, one leading ({\it STEREO-A}) and the other trailing the Earth ({\it STEREO-B}; \citealt{kaiser05}). At the time of ISON observations, {\it STEREO-A} and {\it STEREO-B} were on the far side of the Sun, each $\sim$150$^\circ$ from Earth and $\sim$60$^\circ$ from each other in Carrington longitude. Each spacecraft contains identical instrumentation; we discuss here only the optical imagers in the Sun Earth Connection Coronal and Heliospheric Investigation (SECCHI) instrument package \citep{howard08}. SECCHI contains two heliospheric imagers (HI1, HI2) and two coronagraphs (COR1, COR2). Hereafter we append the {\it STEREO} spacecraft identifier ``A'' or ``B'' to instrument names to refer to specific cameras on specific satellites. Field of view, pixel scale, and bandpass/filters for these telescopes are given in Table~\ref{t:obs_circ}; more detailed information is found in \citet{brueckner95}, \citet{howard08}, \citet{eyles09}, and references therein.

Table~\ref{t:obs_circ} also summarizes the range of dates during which ISON was observed by each telescope. We did not conduct photometric measurements for all images, excluding occasional non-standard configurations and those images in which reliable photometry could not be measured. 
We excluded all HI2 images because the large pixel sizes resulted in nearly every image being contaminated by a background star, all HI1B images because the 180$^\circ$ roll of {\it STEREO-B} required to image ISON resulted in a non-standard background that could not easily be removed, and all COR1 images because background stars are rarely observed, making them difficult to calibrate. As a result, the current work only considers images from C3, C2, HI1A, COR2A, and COR2B. We hope to include additional images in subsequent, more detailed analyses, noting in particular that COR1A was the only telescope to continuously observe ISON without any occultation through perihelion.


\subsection{Reductions}
We used publicly available ``level-0.5'' FITS images as the basis for these analyses. {\it SOHO} images were calibrated as described in \citet{knight10d}.
{\it STEREO} images were processed in an analogous manner using the {\tt secchi\_prep} routine which is part of the SolarSoft IDL package \citep{freeland98}. The exact processing varies slightly by telescope (detailed in \citealt{howard08}) but results in images calibrated to mean solar brightness. 

\begin{figure}
  \centering
  \includegraphics[width=88mm]{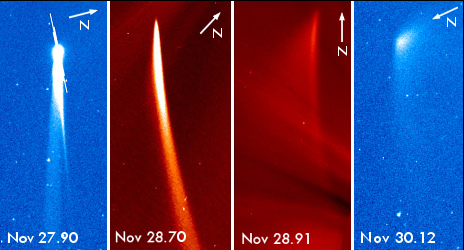}
  \caption[Morphology]{Evolution of ISON's morphology. The first and last images are C3 clear, the middle two images are C2 orange. Images have been resized and reoriented to display as much structure as possible so physical scales are different in each panel. The direction to ecliptic north and the date are displayed on each image, with perihelion occurring Nov 28.779 UT. Color stretches are different in each panel. In the first image, a saturation spike is seen extending from a position angle of $\sim$90$^\circ$ (measured counterclockwise from north), through the central condensation, and out the other side. Image credits: ESA/NASA/SOHO.}
  \label{fig:morphology}
\end{figure}

Our photometric extractions were similar to those in \citet{knight10d}. 
We constructed a median background from nearby images with the identical configuration (binning, filter, polarization, etc.) in which the comet was not within the photometric aperture. 
We centroided on the brightest pixel using a 2-D Gaussian. When no central condensation was evident we estimated the position by eye (see Figure~\ref{fig:morphology}; discussed further below). The flux was measured within circular apertures of radius 350{\arcsec} for HI1, 224{\arcsec} for C3, 103{\arcsec} for COR2, and 71{\arcsec} for C2. Highly different aperture sizes are, unfortunately, necessary due to the varying PSFs by telescope. Due to the large pixel sizes, apertures include nearly all of the coma and some tail. Integrated fluxes were converted to approximate V magnitudes using the coefficients given in \citet{knight10d} for {\it SOHO} and our newly determined coefficients for HI1 and COR2 (similar to those found by \citealt{bewsher10} and \citealt{hui13}). 

ISON's central condensation saturated HI1A images acquired after November 27.63, C2/C3 orange images November 27.13--28.53, and C3 clear images November 27.29--28.46. Saturated HI1A images could not be used for photometric measurements as the processing onboard the spacecraft does not conserve the charge \citep{howard08}. We made photometric measurements in saturated {\it SOHO} images using a routine we developed for C/2011 W3 Lovejoy \citep{knight12c}. Signal loss due to saturation increased as ISON brightened, reaching an estimated few 0.1 mag at peak brightness, but signal loss is similar between consecutive images.
Thus, saturation had minimal effect on the general trends observed in the lightcurve but did flatten the observed slopes of brightening and fading slightly (discussed next).

\section{ANALYSIS}
\label{sec:analysis}
\subsection{Lightcurves}
\begin{figure}
  \centering
  \includegraphics[width=88mm]{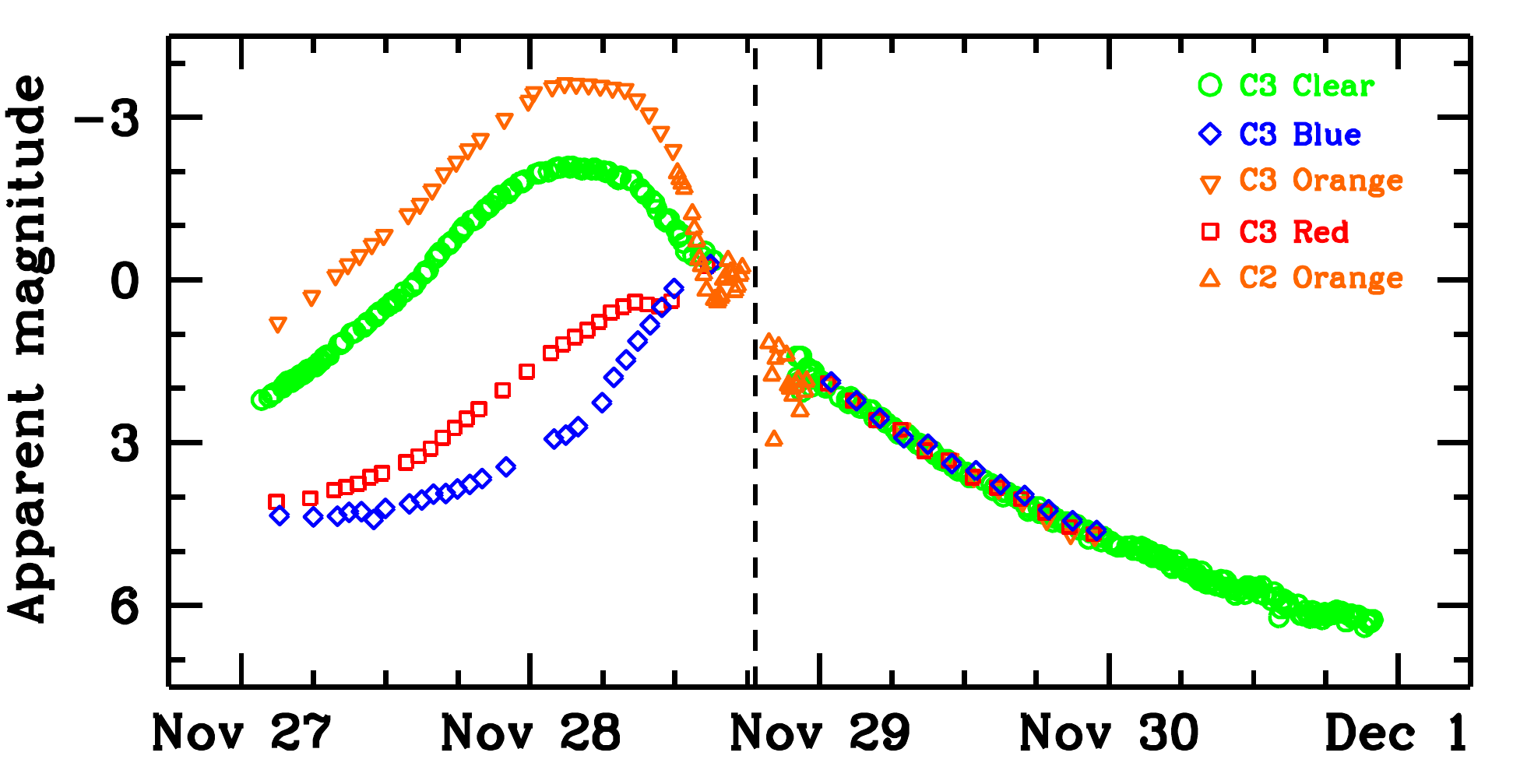}
  \includegraphics[width=88mm]{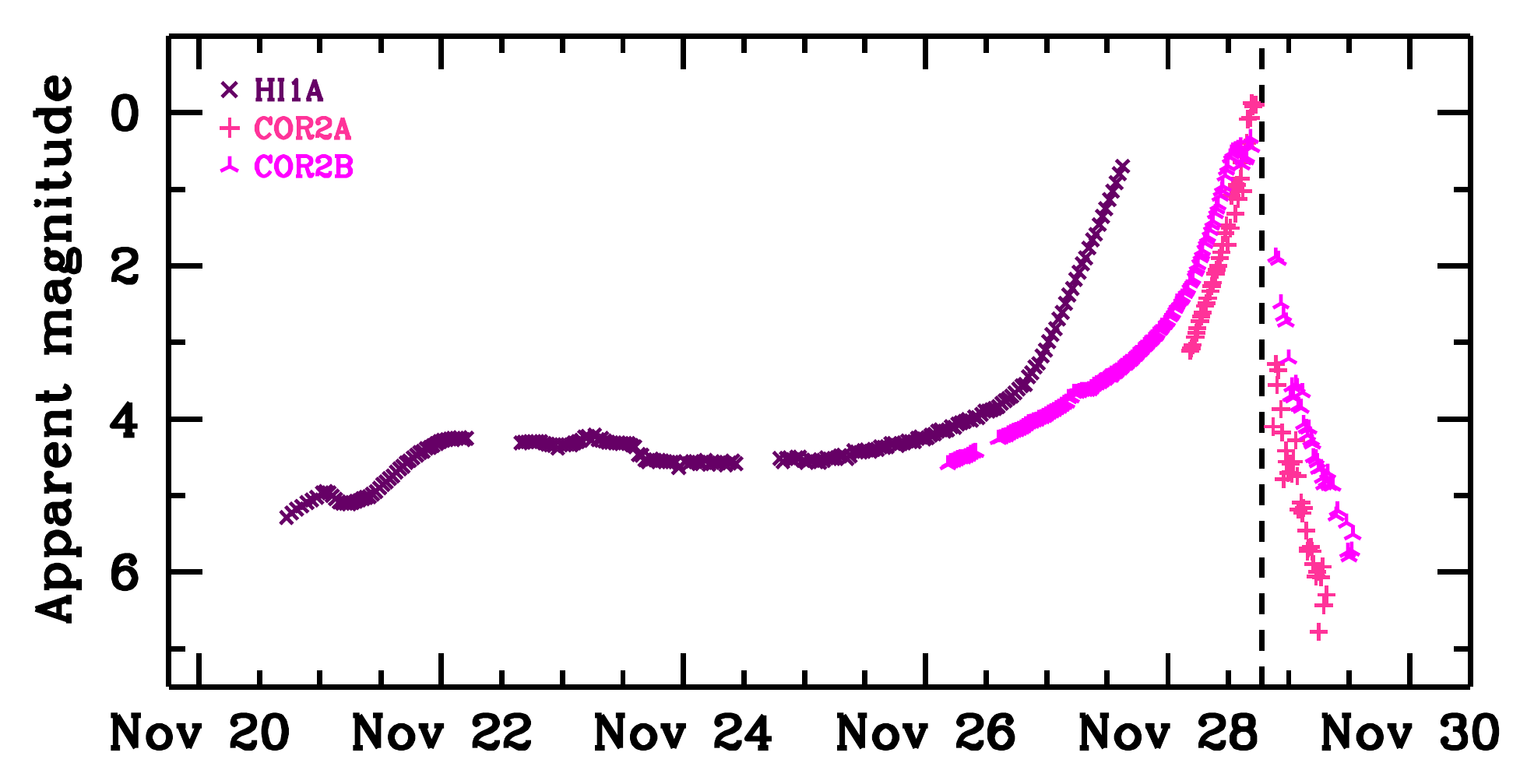}
  \caption[Lightcurve]{Apparent magnitude of ISON as seen by {\it SOHO} (top) and {\it STEREO} (bottom) as a function of time. Symbols are defined in the legend.
Perihelion is denoted by the vertical dashed line. Pre-perihelion C2 magnitudes have been offset by $-$1.5 mag and post-perihelion C2 magnitudes by $-$2.5 to correct for aperture differences between C2 and C3.}
  \label{fig:lc}
\end{figure}

\begin{figure}
  \centering
  \includegraphics[width=88mm]{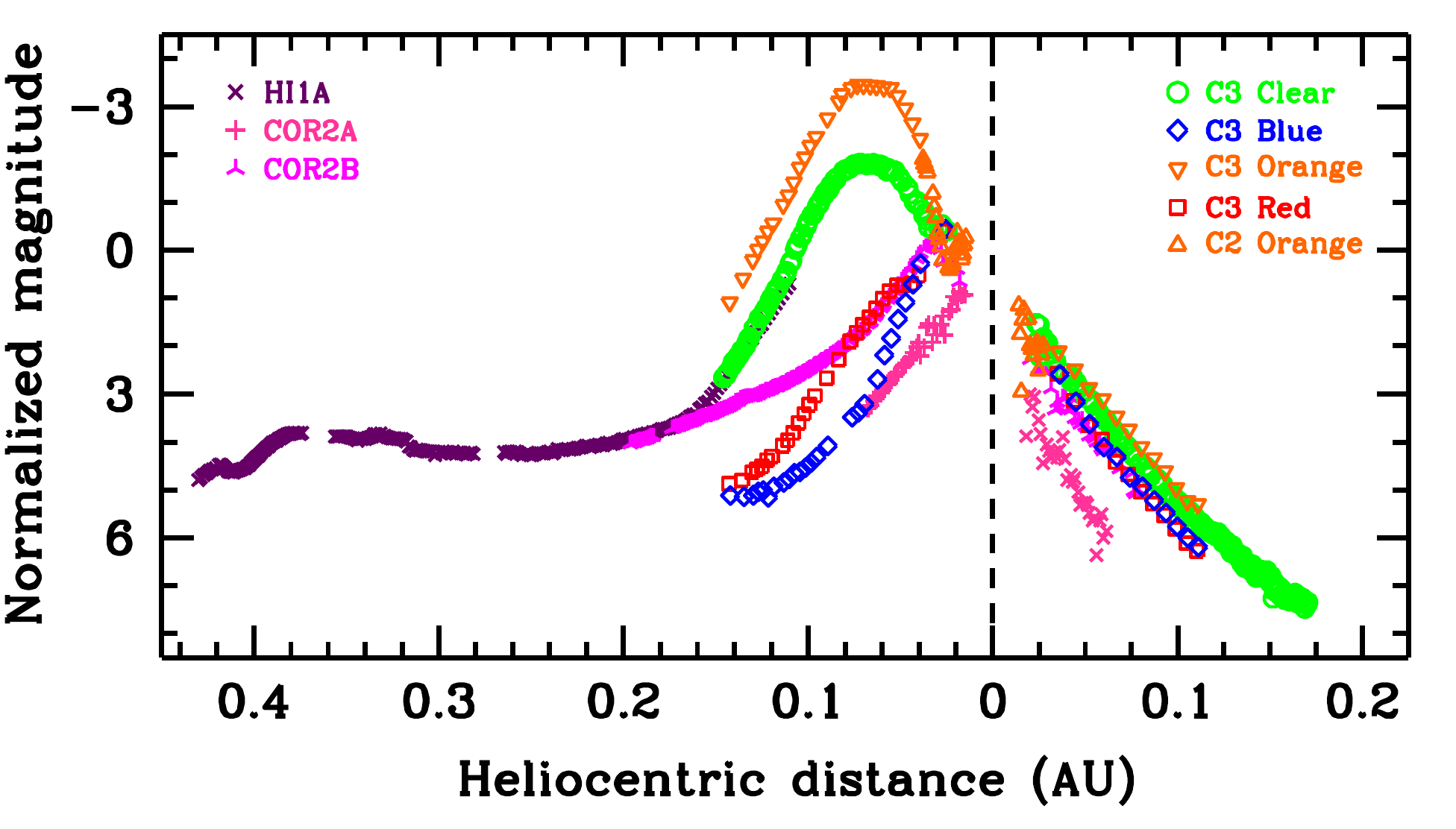}
  \caption[Normalized lightcurve]{Normalized magnitude as a function of heliocentric distance. Magnitudes have been adjusted to a spacecraft-centric distance of 1 AU and a phase angle of 90$^\circ$. All other details are the same as in in Figure~\ref{fig:lc}.}
  \label{fig:norm_lc}
  \label{lastfig}
\end{figure}

We plot ISON's apparent magnitude as a function of time separately for {\it SOHO} and {\it STEREO} in Figure~\ref{fig:lc}; note that these curves have not been adjusted for the differing viewing geometry. We plot the combined dataset normalized to 1 AU from the respective spacecraft and to a phase angle of 90$^\circ$ (following the methodology of \citealt{marcus07b} and \citealt{knight10d}) as a function of heliocentric distance in Figure~\ref{fig:norm_lc}. These normalizations do not change the general trends, but do affect the specific slopes of brightening/fading as well as the magnitude. The phase angle correction makes assumptions about the currently unknown dust-to-gas light ratio in each bandpass, but any changes to the assumptions are likely to have a minimal effect on the interpretations herein.
Henceforth, all analyses refer to the normalized data. Differences in magnitude and shape between filters are primarily due to aperture effects, emission in each bandpass, and differing dust scattering properties than assumed. 
Notably, the aperture sizes were chosen to capture most of the light for typical sungrazing comets \citep{knight10d}. After November $\sim$28.5, ISON had more light in the tail than in the coma so tail contributions in the much larger C3 apertures made these magnitudes much brighter than the smaller C2 apertures. We have applied a correction of $-$1.5 mag (pre-perihelion) and $-$2.5 mag (post-perihelion) to C2 magnitudes to account for this effect.
COR2 magnitudes have not been corrected and therefore appear fainter. Error bars are not plotted as the errors from photon statistics are typically much smaller than the data points.

ISON brightened very gradually November 20--26 while in the HI1A field of view, with an apparent outburst November 21.3--23.5 superimposed. After November 23.5, the brightness fell slightly before leveling off and then brightening ${\propto}r_\mathrm{H}$$^\mathrm{-1.6}$ November 24.8--26.7. Around November 26.7 the slope changed dramatically, brightening ${\propto}r_\mathrm{H}$$^\mathrm{-7.0\ to\ -8.0}$ until November 27.9 in HI1A, C3 orange, and C3 clear. ISON peaked in brightness 
about November 28.0--28.3. It then faded ${\propto}r_\mathrm{H}$$^\mathrm{+2.7\ to\ +6.3}$ in C3 clear, C3 orange, and C2 orange until November 28.6 after which it brightened gradually ${\propto}r_\mathrm{H}$$^\mathrm{-0.8}$ until disappearing behind the occulting disc (C2 orange only). In contrast to the behavior in the clear and orange filters, it never peaked in any other telescope/filter and continued to brighten at various rates (${\propto}r_\mathrm{H}$$^\mathrm{-1.6\ to\ -5.1}$) until disappearing behind the occulting disc of each telescope.

The brightness faded steadily in all telescopes/filters post-perihelion (${\propto}r_\mathrm{H}$$^\mathrm{-2.0\ to\ -2.7}$). The rate of fading was nearly constant in COR2A and COR2B, but became substantially steeper in all four C3 filters after November 29.4 (initially ${\propto}r_\mathrm{H}$$^\mathrm{-0.7\ to\ -1.5}$ then ${\propto}r_\mathrm{H}$$^\mathrm{-2.7\ to\ -3.3}$). ISON was always fainter post-perihelion than at the equivalent heliocentric distance pre-perihelion.

We do not see any evidence for periodic variability in the lightcurve that might be a signature of rotation. This is unsurprising as very small apertures are generally required to extract rotational information for active comets. Such resolution is impossible due to the large pixel scales of {\it SOHO} and {\it STEREO} telescopes.

\subsection{Morphology}
From the time ISON was first visible in HI1A until 2013 November $\sim$28.5 it exhibited a central condensation at its leading edge, with a substantially fainter tail behind it. Abruptly thereafter the strong central condensation disappeared and a broad region of the tail directly behind the former position of the condensation became the brightest area. Until perihelion the leading edge looked ever more elongated without any hint of a central condensation, and the brightest part moved further down the tail. 
For consistency with the rest of the apparition and with our previous studies, 
we continued to measure the flux in the region at the leading edge, though subsequent positions were determined by eye rather than by centroiding.

In the first few hours post-perihelion, a long narrow region roughly tracking the predicted orbit was seen. This broadened and faded; 
specific morphology varied by spacecraft due to viewing geometry. 
At no time was a central condensation seen after perihelion, so all post-perihelion photometry was centered by eye near the leading edge, 
which we assume to approximate the position of any remaining nucleus (per preliminary analyses by \citealt{cbet3731c} and \citealt{cbet3731b}). Photometric measurements were stopped 
when the surface brightness dropped to the point that by-eye centering became erratic.

At no time were structures in the coma evident that might signal fragmentation or correspond to rotation. Although such features were reported from the ground in early to mid-November \citep{cbet3715,cbet3719,cbet3718b}, they were observed at much smaller spatial scales. The extremely large pixels of {\it SOHO} and {\it STEREO} are incapable of capturing this information.


\subsection{Discussion}

The brightness observed in a given bandpass is a combination of reflected solar continuum and emission. Thus, the differing shapes of the lightcurve in different telescopes/filters allows rudimentary compositional information to be gleaned. 
The most valuable comparisons come from the four C3 colored filters. ISON's brightness peaked prior to perihelion in the orange and clear filters but continued to brighten until perihelion in the blue and red filters. The turnover in brightness of Kreutz comets 
has been attributed to the onset of sublimation of refractory grains, notably olivine and pyroxene (e.g., \citealt{kimura02}, \citealt{mann04}). Since the destruction of dust grains should affect all four filters, the lack of a turnover in brightness in either the red or blue filters is puzzling. It may imply that the turnover in the orange and clear filters was not due to sublimation of refractory grains but instead was due to decreasing emission in those bandpasses. Sodium is traditionally assumed to be responsible for the majority of the emission in the orange and clear bandpasses (cf. \citealt{biesecker02}, \citealt{knight10d}) and is therefore the likely culprit. 

Alternatively, there may have been substantial emission in both the red and blue filters that increased faster than the destruction of the dust. 
The only concurrent spectroscopic observations at similar wavelengths of which we are aware were by groups using the McMath-Pierce Solar Telescope (PI: J. Morgenthaler) and the Dunn Solar Telescope (PI. D. Wooden), however, neither group reported obtaining useful data due to weather and ISON's contrast with the background sky. The only previous sungrazing comet with spectroscopic measurements at comparable distances was Ikeya-Seki ($\mathrm{1965f=1965\ S1}$), but no such emissions were detected \citep{preston67,slaughter69}. 
Over similar times, the COR2A and COR2B lightcurves brightened near $r_\mathrm{H}$$^\mathrm{-2}$, suggesting a relatively constant (or even increasing) amount of dust simply responding to the increasing solar flux and, surprisingly, not exhibiting strong evidence of destruction. However, we caution that because all COR2 images are polarized, interpretations are complicated. Clearly, more detailed investigations are needed to properly interpret these behaviors.

The lightcurves were nearly identical in all telescopes/filters post-perihelion suggesting the brightness at this time was dominated by reflected solar continuum off of remnant dust with little to no emission. This is consistent with the idea 
that the nucleus did not survive and therefore there was no active source producing new material. The post-perihelion slope was slightly steeper than the canonical $r_\mathrm{H}^{-2}$ of purely reflected sunlight, and steepened over time.
This was presumably due to dust grains moving out of our photometric aperture without being replenished.

ISON's photometric and morphological behaviors are reminiscent of the small Kreutz comets seen every few days in {\it SOHO} and {\it STEREO} images.
These comets are most often observed in {\it SOHO}'s orange and clear filters, where they brighten rapidly when first observed (${\propto}r_\mathrm{H}^{-7.3\pm2.0}$), peak in brightness at 10--15 $R_\odot$, and fade interior to this with occasional upticks in brightness at $<$7 $R_\odot$ \citep{biesecker02,knight10d}. ISON's slope of brightening from November 26.7--27.9 was nearly identical to the typical slopes of Kreutz comets at similar distances, it peaked in orange/clear at nearly identical distances (12--18 $R_\odot$), and it exhibited a second brightening beginning $\sim$5 $R_\odot$. ISON's orange-clear magnitude difference was generally larger than the typical Kreutz difference of $\sim$1 mag and may indicate compositional or structural differences. Unfortunately, we cannot compare ISON's behavior with typical small Kreutz in the other telescopes/filters as the other {\it SOHO} filters are normally only utilized occasionally and at half resolution, and no systematic study of the Kreutz in {\it STEREO} images has been published (our own work in this is not yet advanced enough for comparison).

While not formally recorded, in our extensive experience with near-Sun comets we can recall numerous instances of the brighter Kreutz comets exhibiting a remarkably similar visual appearance as ISON. When bright enough to assess the morphology, the small Kreutz tend to lose their central condensation about the time the lightcurve peaks, with the leading edge of the comet tapering to a very fine, almost needle-like point, followed by a dense and elongated trail. Thus, ISON's visual appearance in the hours preceding perihelion appears 
entirely consistent with that of a disrupted sungrazing comet.


Typical small Kreutz comets are believed to be tens of meters in radius or smaller and to be destroyed prior to reaching perihelion (cf. \citealt{iseli02,sekanina03,knight10d}). By contrast, Kreutz comet C/2011 W3 Lovejoy, which survived until $\sim$1.6 days post-perihelion despite a considerably smaller perihelion distance of 0.0056 AU, was estimated to be 75--100 m in radius when it disrupted \citep{sekanina12}.  
While we have no reason to believe that ISON's internal structure or makeup were similar to Kreutz comets, applying the same methodology we used to infer Kreutz nucleus sizes \citep{knight10d} suggests the peak brightness was due to the disruption of a $\sim$50 m radius body.


An icy sphere of radius $\sim$50 m is too small to support the published water production rates throughout the apparition (based on the methodology of \citealt{cowan79}). Previous size constraints suggested ISON was $\sim$500 m in radius as of early October (\citealt{cbet3720a}; see also discussion in \citealt{knight13b}). 
As a 
back-of-the-envelope estimate, a crude integration of the SWAN water production rates \citep{iauc9266}, assuming they remain flat from the last measured date (November 23.6) until nearly perihelion, is consistent with only a small remnant being left from a several hundred meter radius object just two months earlier (a similar result was noted by N. Biver\footnote{http://groups.yahoo.com/neo/groups/comets-ml/conversations/topics/22787}). Thus, it is highly likely that ISON was initially substantially larger than 50 m and that nearly all of its mass loss occurred in the days leading up to perihelion.

We now briefly turn our attention to specific features in ISON's lightcurve, noting that deeper analysis is beyond the scope of this letter.
First, the outburst seen in HI1A  November 21.3--23.5 corresponds to the increase in water production noted by \citet{iauc9266} in SWAN data. 
Second, if fragmentation happened after November 20, it most likely occurred around November 21.3 and/or November 26.7, when the lightcurve steepened dramatically. Note, however, that similarly steep brightening to that seen after November 26.7 has been observed in many small Kreutz comets \citep{knight10d} and may be due to rapidly increasing relative sodium emission (J. Marcus, priv. comm.). 
Third, the turnover of the orange/clear brightness November 28.0--28.3 does not necessarily indicate disruption/disintegration at that point. In fact, this was predicted by \citet{cbet3723b} due to the combination of the changing viewing geometry and the destruction of dust grains inside $\sim$0.1 AU (\citealt{mann04} and references therein).


Syndyne and synchrone analyses may eventually constrain the time of final disruption, but the dramatic changes in brightness and morphology began before significant tidal forces should have occurred (cf. \citealt{walsh06,knight13b}). Thus, ISON apparently responded differently to the increasing solar flux than did Lovejoy, which was likely initially smaller yet briefly survived perihelion. This may be tied to natal and/or evolutionary differences between the two comets but will, unfortunately, have to wait to be explored in more detail at a later time.

As we have previously noted, no central condensation was visible in any post-perihelion images. We know that activity from nuclei down to sizes of $\sim$10 m is visible in HI1, and thus we rule out any active surviving component larger than this. The limiting magnitudes of the {\it SOHO} and {\it STEREO} telescopes do not set meaningful upper limits on any surviving inactive fragments; assuming an albedo of 0.04 and an asteroidal phase response, only bare nuclei $\gtrsim$5 km in radius can be excluded. While we consider it highly unlikely that any substantial inactive fragment(s) remains, it cannot be ruled out from this dataset.


\section{SUMMARY}
\label{sec:summary}

The purpose of this Letter has been to quickly convey comet ISON's photometric and morphological behavior in the {\it SOHO} and {\it STEREO} fields of view so that they can be incorporated into the community's narrative of ISON's ultimate demise. Our combined experiences with the small Kreutz comets lead us to draw very obvious parallels between them and ISON's behavior in the hours prior to its perihelion. 
Consequently, we believe that efforts to model ISON's final pre-perihelion moments should consider an object a few tens of meters in diameter.
Furthermore, our investigations suggest that in the weeks and particularly days prior to perihelion, ISON suffered significant mass loss, near complete disruption, and possible devolatilization. 

We anticipate that more detailed analyses of the {\it SOHO} and {\it STEREO} datasets will continue to yield significant insights into ISON's nature and composition. Some topics we hope to explore in future work include modeling of emissions and dust behavior to explain the lightcurve shapes by bandpass, study of the dust properties from polarized images, and 3-D analysis of the evolving coma and tail morphology with time.



\section*{ACKNOWLEDGMENTS}
We are grateful for the {\it SOHO} and {\it STEREO} teams deviating from standard observing plans to optimize collection of ISON data. We thank the anonymous referee for a careful review and Mike A'Hearn, Joe Marcus, and Dave Schleicher for useful discussions. K.B. acknowledges NASA support for the ``Sungrazing Comets Project.''


\begin{thebibliography}{}
\expandafter\ifx\csname natexlab\endcsname\relax\def\natexlab#1{#1}\fi

\bibitem[{{Bewsher} {et~al.}(2010){Bewsher}, {Brown}, {Eyles}, {Kellett},
  {White}, \& {Swinyard}}]{bewsher10}
{Bewsher}, D., {Brown}, D.~S., {Eyles}, C.~J., {et~al.} 2010, \solphys, 264,
  433

\bibitem[{{Biesecker} {et~al.}(2002){Biesecker}, {Lamy}, {St.~Cyr}, {Llebaria},
  \& {Howard}}]{biesecker02}
{Biesecker}, D.~A., {Lamy}, P., {St.~Cyr}, O.~C., {Llebaria}, A., \& {Howard},
  R.~A. 2002, Icarus, 157, 323

\bibitem[{{Boehnhardt} {et~al.}(2013{\natexlab{a}}){Boehnhardt}, {Vincent},
  {Chifu}, {Inhester}, {Oklay}, {Podlipnik}, {Snodgrass}, \&
  {Tubiana}}]{cbet3731c}
{Boehnhardt}, H., {Vincent}, J.~B., {Chifu}, C., {et~al.} 2013{\natexlab{a}},
  Central Bureau Electronic Telegrams, 3731

\bibitem[{{Boehnhardt} {et~al.}(2013{\natexlab{b}}){Boehnhardt}, {Tubiana},
  {Oklay}, {Vincent}, {Hopp}, {Ries}, {Schmidt}, {Riffeser}, \&
  {Goessl}}]{cbet3715}
{Boehnhardt}, H., {Tubiana}, C., {Oklay}, N., {et~al.} 2013{\natexlab{b}},
  Central Bureau Electronic Telegrams, 3715

\bibitem[{{Bonev} {et~al.}(2013){Bonev}, {DiSanti}, {Gibb}, {Villanueva},
  {Paganini}, {Mumma}, \& {Williams}}]{cbet3720b}
{Bonev}, B.~P., {DiSanti}, M.~A., {Gibb}, E.~L., {et~al.} 2013, Central Bureau
  Electronic Telegrams, 3720

\bibitem[{{Brueckner} {et~al.}(1995){Brueckner}, {Howard}, {Koomen},
  {Korendyke}, {Michels}, {Moses}, {Socker}, {Dere}, {Lamy}, {Llebaria},
  {Bout}, {Schwenn}, {Simnett}, {Bedford}, \& {Eyles}}]{brueckner95}
{Brueckner}, G.~E., {Howard}, R.~A., {Koomen}, M.~J., {et~al.} 1995, \solphys,
  162, 357

\bibitem[{{Combi} {et~al.}(2013){Combi}, {Bertaux}, {Quemerais}, {Maekinen}, \&
  {Ferron}}]{iauc9266}
{Combi}, M.~R., {Bertaux}, J.-L., {Quemerais}, E., {Maekinen}, J.~T.~T., \&
  {Ferron}, S. 2013, \iaucirc, 9266

\bibitem[{{Cowan} \& {A'Hearn}(1979)}]{cowan79}
{Cowan}, J.~J., \& {A'Hearn}, M.~F. 1979, Moon and Planets, 21, 155

\bibitem[{{Delamere} {et~al.}(2013){Delamere}, {McEwen}, {Li}, \&
  {Lisse}}]{cbet3720a}
{Delamere}, W.~A., {McEwen}, A.~S., {Li}, J.-Y., \& {Lisse}, C.~M. 2013,
  Central Bureau Electronic Telegrams, 3720

\bibitem[{{Domingo} {et~al.}(1995){Domingo}, {Fleck}, \& {Poland}}]{domingo95}
{Domingo}, V., {Fleck}, B., \& {Poland}, A.~I. 1995, \solphys, 162, 1

\bibitem[{{Eyles} {et~al.}(2009){Eyles}, {Harrison}, {Davis}, {Waltham},
  {Shaughnessy}, {Mapson-Menard}, {Bewsher}, {Crothers}, {Davies}, {Simnett},
  {Howard}, {Moses}, {Newmark}, {Socker}, {Halain}, {Defise}, {Mazy}, \&
  {Rochus}}]{eyles09}
{Eyles}, C.~J., {Harrison}, R.~A., {Davis}, C.~J., {et~al.} 2009, \solphys,
  254, 387

\bibitem[{{Freeland} \& {Handy}(1998)}]{freeland98}
{Freeland}, S.~L., \& {Handy}, B.~N. 1998, \solphys, 182, 497

\bibitem[{{Howard} {et~al.}(2008){Howard}, {Moses}, {Vourlidas}, {Newmark},
  {Socker}, {Plunkett}, {Korendyke}, {Cook}, {Hurley}, {Davila}, {Thompson},
  {St Cyr}, {Mentzell}, {Mehalick}, {Lemen}, {Wuelser}, {Duncan}, {Tarbell},
  {Wolfson}, {Moore}, {Harrison}, {Waltham}, {Lang}, {Davis}, {Eyles},
  {Mapson-Menard}, {Simnett}, {Halain}, {Defise}, {Mazy}, {Rochus}, {Mercier},
  {Ravet}, {Delmotte}, {Auchere}, {Delaboudiniere}, {Bothmer}, {Deutsch},
  {Wang}, {Rich}, {Cooper}, {Stephens}, {Maahs}, {Baugh}, {McMullin}, \&
  {Carter}}]{howard08}
{Howard}, R.~A., {Moses}, J.~D., {Vourlidas}, A., {et~al.} 2008, Space Science
  Reviews, 136, 67

\bibitem[{{Hui}(2013)}]{hui13}
{Hui}, M.-T. 2013, \mnras, 436, 1564

\bibitem[{{Iseli} {et~al.}(2002){Iseli}, {K{\"u}ppers}, {Benz}, \&
  {Bochsler}}]{iseli02}
{Iseli}, M., {K{\"u}ppers}, M., {Benz}, W., \& {Bochsler}, P. 2002, Icarus,
  155, 350

\bibitem[{{Kaiser}(2005)}]{kaiser05}
{Kaiser}, M.~L. 2005, Advances in Space Research, 36, 1483

\bibitem[{{Kimura} {et~al.}(2002){Kimura}, {Mann}, {Biesecker}, \&
  {Jessberger}}]{kimura02}
{Kimura}, H., {Mann}, I., {Biesecker}, D.~A., \& {Jessberger}, E.~K. 2002,
  Icarus, 159, 529

\bibitem[{{Knight} {et~al.}(2010){Knight}, {A'Hearn}, {Biesecker}, {Faury},
  {Hamilton}, {Lamy}, \& {Llebaria}}]{knight10d}
{Knight}, M.~M., {A'Hearn}, M.~F., {Biesecker}, D.~A., {et~al.} 2010, \aj, 139,
  926

\bibitem[{{Knight} \& {Walsh}(2013)}]{knight13b}
{Knight}, M.~M., \& {Walsh}, K.~J. 2013, \apjl, 776, L5 (5pp)

\bibitem[{{Knight} {et~al.}(2012){Knight}, {Kelley}, {Weaver}, {Fernandez},
  {Chesley}, {McNaught}, {Bodewits}, {Lisse}, {Osip}, {Dello Russo}, \&
  {Battams}}]{knight12c}
{Knight}, M.~M., {Kelley}, M.~S., {Weaver}, H.~A., {et~al.} 2012, in
  AAS/Division for Planetary Sciences Meeting Abstracts, Vol.~44, \#514.02

\bibitem[{{Mann} {et~al.}(2004){Mann}, {Kimura}, {Biesecker}, {Tsurutani},
  {Gr{\"u}n}, {McKibben}, {Liou}, {MacQueen}, {Mukai}, {Guhathakurta}, \&
  {Lamy}}]{mann04}
{Mann}, I., {Kimura}, H., {Biesecker}, D.~A., {et~al.} 2004, Space Science
  Reviews, 110, 269

\bibitem[{{Marcus}(2013)}]{cbet3723b}
{Marcus}, J. 2013, Central Bureau Electronic Telegrams, 3723

\bibitem[{{Marcus}(2007)}]{marcus07b}
{Marcus}, J.~N. 2007, Int. Comet Quart., 39

\bibitem[{{Opitom} {et~al.}(2013){Opitom}, {Jehin}, {Manfroid}, \&
  {Gillon}}]{cbet3719}
{Opitom}, C., {Jehin}, E., {Manfroid}, J., \& {Gillon}, M. 2013, Central Bureau
  Electronic Telegrams, 3719

\bibitem[{{Preston}(1967)}]{preston67}
{Preston}, G.~W. 1967, \apj, 147, 718

\bibitem[{{Sekanina}(2003)}]{sekanina03}
{Sekanina}, Z. 2003, \apj, 597, 1237

\bibitem[{{Sekanina}(2013)}]{cbet3731b}
---. 2013, Central Bureau Electronic Telegrams, 3731

\bibitem[{{Sekanina} \& {Chodas}(2012)}]{sekanina12}
{Sekanina}, Z., \& {Chodas}, P.~W. 2012, \apj, 757, 127

\bibitem[{{Slaughter}(1969)}]{slaughter69}
{Slaughter}, C.~D. 1969, \aj, 74, 929

\bibitem[{{Walsh} \& {Richardson}(2006)}]{walsh06}
{Walsh}, K.~J., \& {Richardson}, D.~C. 2006, \icarus, 180, 201

\bibitem[{{Ye} {et~al.}(2013){Ye}, {Hui}, \& {Gao}}]{cbet3718b}
{Ye}, Q., {Hui}, M.~T., \& {Gao}, X. 2013, Central Bureau Electronic Telegrams,
  3718

\end{thebibliography}

\label{lastpage}

\end{singlespace}


\renewcommand{\baselinestretch}{0.78}
\renewcommand{\arraystretch}{1.0}

\begin{deluxetable}{lccccccc}  
\tabletypesize{\scriptsize}
\tablecolumns{8}
\tablewidth{0pt} 
\setlength{\tabcolsep}{0.03in}
\tablecaption{Summary of observations}
\tablehead{   
  \colhead{Spacecraft}&
  \colhead{Telescope}&
  \colhead{Field of}&
  \colhead{Pixel Scale}&
  \colhead{Bandpass}&
  \colhead{Pre-perihelion}&
  \colhead{Post-perihelion}&
  \colhead{Used In}\\
  \colhead{}&
  \colhead{}&
  \colhead{View\tablenotemark{a} ($^\circ$)}&
  \colhead{({\arcsec} pix$^{-1}$)}&
  \colhead{(\AA)}&
  \colhead{Image Range}&
  \colhead{Image Range}&
  \colhead{Analyses?}
}
\startdata
{\it SOHO}&C3&1.0--8.0&56.0&C,B,O,R\tablenotemark{b}&Nov 27--28&Nov 28--30&Yes\\
{\it SOHO}&C2&0.3--1.6&11.9&B,O,R\tablenotemark{b}&Nov 28&Nov 28&Yes\\
{\it STEREO-A}&HI2&20.7--90.7&240.0&4000--10000&Oct 10--Nov 23&Unobserved&No\\
{\it STEREO-A}&HI1&4.0--24.0&70.0&6300--7300\tablenotemark{c}&Nov 20--28&Dec 1--7&Only pre-perihelion\\
{\it STEREO-A}&COR2&0.5--4.0&14.7&6500--7500\tablenotemark{d}&Nov 28&Nov 28--30&Yes\\
{\it STEREO-A}&COR1&0.4--1.0&3.8&6500--6600\tablenotemark{d}&Nov 28&Nov 28&No\\
{\it STEREO-B}&HI1&4.0--24.0&70.0&6300--7300\tablenotemark{c}&Oct 24--Nov 25\tablenotemark{e}&Unobserved&No\\
{\it STEREO-B}&COR2&0.5--4.0&14.7&6500--7500\tablenotemark{d}&Nov 26--28&Nov 28--30&Yes\\
{\it STEREO-B}&COR1&0.4--1.0&3.8&6500--6600\tablenotemark{d}&Nov 28&Nov 28&No\\
\enddata
\tablenotetext{a} {Annular fields of view centered on the Sun except for HI1 and HI2 which are offset by 14.0$^\circ$ and 55.7$^\circ$, respectively, along the Earth-Sun line as viewed by the spacecraft.}
\tablenotetext{b} {$\mathrm{C=Clear}$ (4000--8500\AA), $\mathrm{B=Blue}$ (4200--5200\AA), $\mathrm{O=Orange}$ (5400--6400\AA), $\mathrm{R=Red}$ (7300--8350\AA)}
\tablenotetext{c} {Also has significant blue transmission \citep{bewsher10}.}
\tablenotetext{d} {All COR1 and COR2 images are polarized; we use only ``total polarization'' COR2 images.}
\tablenotetext{e} {ISON only appears in images rolled by 180$^{\circ}$ and acquired intermittently.}

\label{t:obs_circ}
\label{lasttable}
\end{deluxetable}

\renewcommand{\baselinestretch}{0.8}


\end{document}